\documentclass[letterpaper, 10pt, conference]{ieeeconf}
\IEEEoverridecommandlockouts 
\usepackage[latin1]{inputenc}
\usepackage{amsmath, amsfonts, amssymb, array}

\newtheorem{thm}{Theorem}[section]
\newtheorem{prop}[thm]{Proposition}
\newtheorem{lemma}[thm]{Lemma}
\newtheorem{cor}[thm]{Corollary}

\newtheorem{definition}[thm]{Definition}
\newtheorem{example}[thm]{Example}

\title{\LARGE \bf New Improvements on the Echelon-Ferrers Construction}
\author{Anna-Lena Trautmann and Joachim Rosenthal $^*$
\thanks{$^*$ This research was partially supported by Swiss National Science
    Foundation under Grant no. 126948 .}
\thanks{The authours are with the Institute of Mathematics, University of
Zurich, Winterthurerstrasse 190, 8057 Zurich, Switzerland,
        {\tt\small www.math.uzh.ch/aa} .}%
}

\begin{document}

\maketitle
\thispagestyle{empty}
\pagestyle{empty}

\begin{abstract}
  We show how to improve the echelon-Ferrers construction of random network
codes introduced in \cite{et08u} to attain codes of larger size for a given
minimum distance.
\end{abstract}

\vspace{0.5cm}



\section{Preliminaries}

In network coding one is looking at the transmission of information
through a directed graph with possibly several senders and several
receivers. One can increase the throughput by doing linear combinations at
intermediate nodes of the network.  If the underlying
topology of the network is unknown we speak about \textit{random network
coding}. Since linear spaces are invariant under linear combinations, they are
exactly what is needed as codewords (see \cite{ko08}). It is helpful (e.g. for
decoding) to constrain oneself to subspaces of a fixed
dimension, in which case we talk about \emph{constant dimension codes}. 

Let $\mathbb{F}_q$ be the finite field with $q$ elements (where
$q=p^r$ and $p$ prime). The \textit{projective space}
$\mathbb{P}^{n-1}$ of order $n-1$ over $\mathbb{F}_q$ is the set of
all 1-dimensional subspaces of $\mathbb{F}_q^n$. The set of all subspaces of
$\mathbb{F}_q^n$ of
dimension $k$ is called \emph{Grassmannian}, denoted by
$\mathcal{G}(k,n)$.

It is a well-known result that 
\[|\mathcal{G}(k,n)|=\left[\begin{array}{c}n\\k\end{array}\right]_q :=
\prod_{i=0}^{k-1} \frac{q^{n-i}-1}{q^{k-i}-1}\]

Let $U\in Mat_{k\times n}(\mathbb{F}_q)$ be a matrix such that
$\mathcal{U}= \mathrm{row space}(U)$. The matrix $U$ is usually not
unique. Indeed one can notice that
\[\mathcal{U}=\mathrm{row space}(U)= \mathrm{row space}(A\cdot U)\] 
for any $A\in GL_k(\mathbb{F}_q)$, i.e. any $k$-dimensional subspace is
stable under the action of $GL_k(\mathbb{F}_q)$.
However there exists a unique matrix representation of
elements of the Grassmannian, namely the reduced row echelon forms.

The \emph{subspace distance} is a metric on $\mathcal{G}(k,
n)$ given by
\begin{align*}
d_S(\mathcal{U},\mathcal{V}) =& 2(k - \dim(\mathcal{U}\cap
\mathcal{V}))\\
=& 2\cdot rank\left[ \begin{array}{c} U \\ V \end{array}
\right] - 2k
\end{align*}
for any $\mathcal{U},\mathcal{V} \in
\mathcal{G}(k, n)$.

A constant dimension code $C$ is simply a subset of the Grassmannian
$\mathcal{G}(k,n)$. If the distance between any two elements of it is
greater than or equal to $2\delta$ we say that $C$ has minimum
distance $2\delta$ and call it a $[n, 2\delta, |C|, k]$-code.

\textit{Remark:}
$A[n,2\delta,k]$ denotes the maximal cardinality of a code in $\mathcal{G}(k,n)$
with minimum distance $2\delta$. It holds that
$A[n,2\delta,k]=A[n-k,2\delta,k]$ (orthogonal complement \cite{et08p}). Therefore we
restrict our studies to the case $2k\leq n$.


In classical coding theory over $\mathbb{F}_2$ the \emph{Hamming distance} $d_H$
between two vectors of the same length is defined to be the number of positions
in which they differ. \emph{Lexicodes}, also called lexicographic codes \cite{co86a5}, are greedily generated codes with minimum
distance $d$, where one starts with the first element in lexicographic order and
adds the lexicographic next element that fulfills the distance requirement.

In the space of $m\times n$-matrices over $\mathbb{F}_q$ the \emph{rank distance}
between two elements $X$ and $Y$ is defined to be
\[d_R(X,Y):= rank(X-Y)\]



In \cite{et08u} T. Etzion and N. Silberstein introduced the Echelon-Ferrers
construction for which we need the following definitions:

\begin{definition}
\begin{enumerate}
 \item
The \emph{identifying vector} $v(U)$ of a matrix $U$ in reduced row echelon form
is the binary vector of length $n$ and weight $k$ such that the $1$'s of $v(U)$
are in the positions where $U$ has its pivots (also called leading ones).
 \item 
A \emph{Ferrers diagram} $F$ is a pattern of dots such that all dots are
shifted to the right of the diagram and the number of dots in a row is less than or
equal to the number of dots in the row above.
 \item 
A \emph{Ferrers diagram code} $C_F$ is a rank-metric code such that all entries
not in the Ferrers diagram $F$ are $0$.
\end{enumerate}
\end{definition}

The echelon-Ferrers code construction is a multilevel construction:
First we construct the \emph{skeleton code} by choosing a binary linear code of
length $n$, weight $k$ and minimum Hamming distance $\delta$ and finding the
corresponding matrices such that these code words are their identifying vectors.

Then we fill each of the originated Ferrers diagrams with a compatible Ferrers
diagram code with minimum rank distance $\delta$. 

One can easily check (with the following propositions) that the row spaces of the above constructed matrices form
a constant dimension code in $\mathcal{G}(k,n)$ with minimum subspace distance
$2\delta$.

\textit{Remark:}
The set of all reduced row echelon forms with the same identifying vector is
exactly a Schubert cell.

\begin{prop}
Let $U$ and $V$ be in the same Schubert cell, i.e. $v(U)=v(V)$. Then
\[d_S(\mathcal{U},\mathcal{V})= d_R((C_F)_U,(C_F)_V)\]
where $(C_F)_U$ and $(C_F)_V$ denote the submatrices of $U$ and $V$,
respectively, without the columns of their pivots.
\end{prop}

\begin{prop}\cite{et08p1}
Let $\mathcal{U}$ and $\mathcal{V} \in \mathcal{G}(k,n)$ and $U$ and $V$ their
representation matrices, respectively. Then
\[d_S(\mathcal{U},\mathcal{V})\geq d_H(v(U),v(V))\]
\end{prop}

\textit{Remark:}
It is a hard problem to understand which skeleton code leads to the largest subspace code. Although lexicodes themselves are not among the largest binary linear codes they are a good choice for skeleton codes.

\begin{example}
We want to construct a code in $\mathcal{G}(3,6)$ with minimum distance $4$,
hence we start with the binary lexicode of length $6$, weight $3$ and distance
$2$ as skeleton code. This code has the following three codewords:
\[(111000), (100110), (010101)\]
The corresponding echelon-Ferrers forms are:
\[ \left(\begin{array}{cccccc}1 & 0 & 0 & \bullet & \bullet &\bullet \\0
& 1 & 0 & \bullet& \bullet & \bullet \\0 & 0 & 1 & \bullet&\bullet &\bullet
\end{array}\right), 
\left(\begin{array}{cccccc}1 & \bullet & \bullet & 0 & 0 &\bullet\\0 & 0 & 0 & 1
& 0 &\bullet \\0 & 0 & 0 & 0 & 1&\bullet  \end{array}\right), \]
\[\left(\begin{array}{cccccc}0 & 1 & \bullet & 0 & \bullet & 0 \\0 & 0 & 0 & 1 &
\bullet & 0 \\0 & 0 & 0 & 0 & 0 & 1\end{array}\right)\]
We can fill the Ferrers diagrams with rank distance codes of size $q^6, q^2$ and
$q$, respectively. Thus we constructed a $[6,4,q^6+q^2+q,3]$-code.
\end{example}

The following theorem was stated and proved in \cite{et08p1}.
\begin{thm}
Let $F$ be a Ferrers diagram and $C_F$ the corresponding Ferrers diagram code.
Then
\[|C_F| \leq q^{\min_i\{w_i\}}\]
where $w_i$ is the number of dots in $F$ which are not contained in the first
$i$ rows and the rightmost $\delta-1-i$ columns ($0\leq i\leq \delta-1$). 
Moreover the bound can be obtained for (at least) $\delta=1,2$.
\end{thm}

For certain Ferrers diagrams this gives us a nice formula on the size of the
Ferrers diagram code.
\begin{cor}
Let $a\geq b$ and $F$ be an $a\times b$ Ferrers diagram. 
Assume that each one of the rightmost $\delta -1$ columns of $F$ has $a$ dots.
Then
\[\dim C_F = \sum_{i=1}^{b-\delta+1} \gamma_i\]
where $\gamma_i$ is the number of dots in the $i$-th column of $F$.

Similarly let $a\leq b$ and $F$ be an $a\times b$ Ferrers diagram. 
Assume that each of the first $\delta -1$ rows of $F$ has $a$ dots. Then
\[\dim C_F = \sum_{i=\delta-1}^{b} \hat{\gamma}_i\]
where $\hat{\gamma}_i$ is the number of dots in the $i$-th row of $F$.
\end{cor}


\section{Improvement on the Packing}

Some skeleton code words lead to a Ferrers diagram where one can remove dots and
still achieve the same size of the corresponding Ferrers diagram code.  
We can improve the size of our subspace codes if we take these removable dots
into account.



\begin{example}\label{fer1}
All of the following Ferrers diagrams give rise to a Ferrers
diagram code with minimum distance $4$ of size $q^3$, since the minimum number of dots not contained
either in the first row or in the last column is $3$.
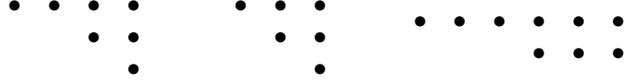
\[ \left.\begin{array}{cccc}\bullet & \bullet & \bullet & \bullet \\ &  &
\bullet & \bullet \\ &  &  & \bullet\end{array}\right.  \hspace{0.8cm} 
\begin{array}{ccc}\bullet & \bullet & \bullet \\ &  \bullet& \bullet \\ &  &
\bullet\end{array} \hspace{0.8cm}
\begin{array}{cccccc}\bullet & \bullet & \bullet & \bullet & \bullet & \bullet
\\  &  &   & \bullet  & \bullet & \bullet\end{array}\]
\end{example}

\vspace{0.1cm}

\begin{definition}
Let $F$ be a Ferrers diagram and $f_{ij}$ be the dot in the $i$-th row and
$j$-th column from the right. $F\backslash f_{ij}$ denotes the Ferrers diagram
$F$ after removing $f_{ij}$. We call a set of dots $\{f_{ij}\}$ \emph{pending}
if they are in the first row and the leftmost columns of the Ferrers diagram and
\[|C_F| = |C_{F\backslash \{f_{ij}\}}|\] 
\end{definition}

\textit{Remark:}
One can also define pending dots in the rightmost column on the very bottom and
translate the following results to that setting. 

\begin{example}
In Example \ref{fer1} the first and the second Ferrers diagrams lead to the
same-size rank metric code. Thus the top leftmost dot of the left diagram is
pending.
\end{example}



\begin{prop}
Let $z_i$ be the number of $0$'s after the $i$-th $1$ of the identifying vector and
\[\mathfrak{p}:=\sum_{i=1}^k z_i - \max\limits_{z_l\neq 0}l = n-k-z_0-\max\limits_{z_l\neq
0}l .\] 
Then the following holds:
\begin{enumerate}
 \item 
If $z_i=0$ for all but one $i>1$ then $\mathfrak{p}=0$. If $z_i=0$ for all $i>1$, i.e. $z_1=n-k-z_0$, then $\mathfrak{p}=z_1$.
 \item 
If $\mathfrak{p} > 0$ then there are $\min\{\mathfrak{p}, z_1\} $ pending dots in the top row of the Ferrers diagram.
\end{enumerate}

\begin{proof}
\begin{enumerate}
 \item 
If $z_i=0$ for all but one $i$, i.e. all $0$'s are in one block, then the
Ferrers diagram is a rectangle, hence there are no pending dots (except if it is
a line, then the whole diagram is pending).  
 \item
The number of dots not located in the last column is 
\(\sum_{i=2}^k i\cdot z_i - \max_{z_l\neq 0} l\)
the number of dots not located in the first row is 
\(\sum_{i=2}^k (i-1)\cdot z_i .\)
Thus the number of dots without importance for the Ferrers diagram code is the difference of the two: 
\[\sum_{i=1}^k i \cdot z_i - \max_{z_l\neq 0}l - \sum_{i=2}^k (i-1)\cdot z_i =
\sum_{i=1}^k z_i - \max\limits_{z_l\neq 0}l\]
Pending dots can only occur in the first row, 
hence their number cannot be larger than $z_1$. 
\end{enumerate}
\end{proof}

\end{prop}


\begin{thm}
Let $v(U)$ be an identifying vector of length $n$ and constant weight $k$ such
that the corresponding Ferrers diagram has a set of pending dots in the first
row. Let $v(V)$ be another identifying vector of the same length and weight
(subsequent in lexicographic order) such that the first $1$ is in the same
position as for $v(U)$ and $d_H(v(U),v(V))=2\delta-2$. Fix the matrix entries at
the positions of the pending dots as a $\mathfrak{p}$-tuple $\mu$ for all elements of the
cell of $v(U)$ and as a $\mathfrak{p}$-tuple $\nu \neq \mu$ for all elements of the cell of
$v(V)$. Then
\[rank\left[\begin{array}{c} U \\ V\end{array}\right] \geq k + \delta\]
for any $U_i$ in the cell of $v(U)$ and $V_j$ in the cell of $v(V)$.

\begin{proof}
From the Hamming distance of the identifying vectors we know that 
\[rank\left[\begin{array}{c} U \\ V\end{array}\right]\geq
k+\delta -1 . \]
Moreover the first rows of $U$ and $V$ are linearly independent since $\mu \neq
\nu$. Together with the fact that all other leading ones appear to the right of
$\mu$ and $\nu$, this proves the statement.
\end{proof}

\end{thm}

\begin{cor}
Let $v(U)$ and $v(V)$ as before and fill the Ferrers diagrams of $v(U)$ and
$v(V)$ with a respective Ferrers diagram code of minimum distance $\delta$. The
corresponding row spaces of this set of matrices is a constant dimension code in
$\mathcal{G}(k,n)$ with minimum distance $2\delta$.

\begin{proof}
One knows already that $d_S(\mathcal{U}_i,\mathcal{U}_j)=2\delta $ and
$d_S(\mathcal{V}_i,\mathcal{V}_j)=2\delta $, hence inside the cell the minimum
distance is out of question. Because of the above theorem we know that 
\[d_S(\mathcal{U}_i, \mathcal{V}_j) = 2\cdot  rank \left[\begin{array}{c} U_i \\ V_j\end{array}\right]- 2k \geq 2\delta
\]
\end{proof}
\end{cor}

\begin{example}\label{ex1}
Let us consider the skeleton code word $(1001100)$, thus our cell is of the type
\[ \left(\begin{array}{ccccccc}1 & \fbox{$\bullet$} & \bullet & 0 & 0
&\bullet&\bullet \\0 & 0 & 0 & 1 & 0 &\bullet &\bullet \\0 & 0 & 0 & 0 &
1&\bullet&\bullet  \end{array}\right)\]
where the dot in the box marks the position of the pending dot. 
We choose $(1000110)$ as second skeleton code word and fix the pending position
as $0$ in the first cell and as $1$ in the second: 
\[ \left(\begin{array}{ccccccc}1 & \fbox{0} & \bullet & 0 & 0
&\bullet&\bullet \\0 & 0 & 0 & 1 & 0 &\bullet &\bullet \\0 & 0 & 0 & 0 &
1&\bullet&\bullet  \end{array}\right) \] 
\[\left(\begin{array}{ccccccc}1 & \fbox{1} & \bullet &0& \bullet
&0&\bullet \\0 & 0 & 0 & 1 & \bullet &0&\bullet \\0 & 0 & 0 & 0 & 0&1&\bullet 
\end{array}\right)\]
Although the Hamming distance between the two identifying vectors is $2$ we
obtain a subspace distance of $4$.
\end{example}


\section{Improved Code Construction}

We will now explain the new construction:
\begin{enumerate}
 \item Begin the skeleton code with the first lexicode element $(11...10...0)$ and
fill the echelon-Ferrers form with a maximum rank distance code.
 \item Choose the second skeleton code word as the next lexicode element and fix
the set of pending dots (if there are any) of the Ferrers diagram as
$\mu_1$. Fill the echelon-Ferrers form with a Ferrers diagram code.
 \item For the next skeleton code word choose the first $1$ in the same
positions as before and use the next lexicode element of distance $\geq 2\delta
-2$ from the other elements with the same pending dots and $\geq 2\delta $ from
any other skeleton code word. Fix the pending dots as a tuple $\mu_i$ different from the
tuples already used for echelon-Ferrers forms where the Hamming distance of the
identifying vectors is $2\delta -2$. Fill the echelon-Ferrers form with a
Ferrers diagram code.
 \item Repeat step 3 until no possibilities for a new skeleton code word with
the fixed $1$ are left.
 \item In the skeleton code choose the next vector in lexicographic order that
has distance $\geq 2\delta$ from all other skeleton code words and repeat steps
2,3 and 4.
\end{enumerate}

\begin{prop}
Let us consider the above construction and $\delta = k$. Then every originating Ferrers diagram is of rectangular shape and has no pending dots.

\begin{proof}
 Since the first skeleton code word has all $1$'s in a block, there are no pending dots. Because of the minimum distance the second skeleton code word is 
 \[(\underbrace{0\hdots 0}_k \underbrace{1\hdots 1}_k 0\hdots 0)\]
 thus there are again no pending dots. The same argument holds for all following code words.
\end{proof}

\end{prop}

It follows that for codes of maximal distance, i.e.  $2\delta = 2k$, the construction is exactly the classical echelon-Ferrers construction.

\begin{lemma}
The first skeleton code word $(1...10...0)$ always leads to a component code of
size $q^{(n-k)(k-\delta+1)}$.
\end{lemma}

For the remain of this section we look at the case $\delta=2$. Thus $\dim C_F$
is equal to the minimum number of dots that are either not in the first row or
the last column of a Ferrers diagram $F$. 

\begin{example}
We want to construct a code in $\mathcal{G}(3,7)$ with minimum distance $4$.
\begin{enumerate}
 \item We choose the first skeleton code word $1110000$, whose echelon-Ferrers
form can be filled with a maxi\-mum rank distance code of size $q^{8}$.
 \item \begin{enumerate}
	\item The second skeleton code word $1001100$ leads to a Ferrers diagram
with one pending dot (see example \ref{ex1}).
	\item Fix the pending dot as $0$.
	\item The echelon-Ferrers form can be filled with a Ferrers diagram code
of size $q^4$.
	\end{enumerate}
\[ \left(\begin{array}{ccccccc}1 & \fbox{0} & \bullet & 0 & 0 &\bullet&\bullet \\0 & 0
& 0 & 1 & 0 &\bullet &\bullet \\0 & 0 & 0 & 0 & 1&\bullet&\bullet 
\end{array}\right)\]

 \item \begin{enumerate}
	\item The next skeleton code word $1001010$ leads to a Ferrers diagram
with a pending dot in the same position as before.
	\item Fix the pending dot as $1$.
	\item The echelon-Ferrers form can be filled with a Ferrers diagram code
of size $q^3$.
	\end{enumerate}
\[ \left(\begin{array}{ccccccc}1 & \fbox{1} & \bullet & 0&\bullet & 0 &\bullet \\0 & 0
& 0 & 1&\bullet  & 0 &\bullet \\0 & 0 & 0 & 0 & 0 & 1&\bullet 
\end{array}\right)\]

 \item \begin{enumerate}
	\item The next skeleton code word $1000101$ leads to a Ferrers diagram
with a pending dot in the same position as before. (Actually there are two
pending dots but we can only make use of the one from before.)
	\item Fix the pending dot as $1$.
	\item The echelon-Ferrers form can be filled with a Ferrers diagram code
of size $q$.
	\end{enumerate}
\[ \left(\begin{array}{ccccccc}1 & \fbox{1} & \bullet & \bullet & 0 &\bullet&0 \\0 & 0
& 0 & 0 & 1 &\bullet &0 \\0 & 0 & 0 & 0 & 0 & 0 & 1  \end{array}\right)\]

 \item The following skeleton code word $0101001$ leads to a echelon-Ferrers
form that can be filled with a Ferrers diagram code of size $q^2$.
 \item In analogy $0100110$ can be filled with a Ferrers diagram code of size
$q^2$.
 \item The last skeleton code word $0010011$ can be filled with a Ferrers
diagram code of size $1$.
\end{enumerate}
Hence we constructed a $[7,4,q^8+q^4+q^3+2q^2+q+1,3]$-code, which is larger
than the code constructed by the standard echelon-Ferrers construction.
\end{example}

The following tables show some examples where the new construction leads to larger codes than the one before. All codes have minimum distance $4$.

\begin{center}
\begin{tabular}{|l|l|p{5.5cm}|}
\hline 
n  & k & classical echelon-Ferrers construction \\ \hline 
7 & 3 & $q^8+q^4+q^3+q^2+2q+1$ \\
8 & 3 & $q^{10}+q^6+q^5+2q^4+q^3+q^2$ \\
9 & 3 & $q^{12}+q^8+q^7+2q^6+q^5+q^4+1$ \\
\hline
\end{tabular}

\vspace{0.5cm}

\begin{tabular}{|l|l|p{5.5cm}|}
\hline 
n  & k & new echelon-Ferrers construction \\ \hline 
7 & 3 &  $q^8+q^4+q^3+2q^2+q+1$\\
8 & 3 & 
$q^{10}+q^6+q^5+2q^4+2q^3+2q^2+q+1$\\
9 & 3 & 
$q^{12}+q^8+q^7+2q^6+2q^5+3q^4+2q^3+$ \\ &&$2q^2+q+1$\\
\hline
\end{tabular}
\end{center}


\section{Conclusion and Open Problems}

In this work we show how the echelon-Ferrers construction by T. Etzion and N. Silberstein can be improved by con\-si\-dering the pending dots of the obtained Ferrers diagrams. We show when and how many pending dots occur depending on the underlying identifying vector. In the end some examples of code sizes were given, which are larger than codes obtained by other constructions in \cite{et08p1,ma08p} and \cite{si08a}. Although over $\mathbb{F}_2$ some larger codes have been found in \cite{ko08p}, some of our codes are the largest codes found so far in the general setting over $\mathbb{F}_q$.
 
Since in this paper we only considered pending dots in the top row, an open problem
is to look at a generalized setting where a set of pending dots can occur in
the top rows (more than one). Moreover one could investigate if improvements can be made by looking at pending dots in the top row as well as in the rightmost column.


\bibliography{/home/a/rosen/Bib/huge.bib}

\def\cprime{$'$} \def\polhk#1{\setbox0=\hbox{#1}{\ooalign{\hidewidth
  \lower1.5ex\hbox{`}\hidewidth\crcr\unhbox0}}}
  \def\polhk#1{\setbox0=\hbox{#1}{\ooalign{\hidewidth
  \lower1.5ex\hbox{`}\hidewidth\crcr\unhbox0}}} \def\cprime{$'$}
  \def\cprime{$'$} \def\cprime{$'$} \def\cprime{$'$}
\begin{thebibliography}{1}

\bibitem{co86a5}
John~H. Conway and N.~J.~A. Sloane.
\newblock Lexicographic codes: error-correcting codes from game theory.
\newblock {\em IEEE Trans. Inform. Theory}, 32(3):337--348, 1986.

\bibitem{et08p1}
T.~Etzion and N.~Silberstein.
\newblock Construction of error-correcting codes for random network coding.
\newblock In {\em Electrical and Electronics Engineers in Israel, 2008. IEEEI
  2008. IEEE 25th Convention of}, pages 070--074, Dec. 2008.

\bibitem{et08u}
T.~Etzion and N.~Silberstein.
\newblock Error-correcting codes in projective spaces via rank-metric codes and
  {F}errers diagrams.
\newblock arXiv:0807.4846, 2008.

\bibitem{et08p}
T.~Etzion and A.~Vardy.
\newblock Error-correcting codes in projective space.
\newblock In {\em Information Theory, 2008. ISIT 2008. IEEE International
  Symposium on}, pages 871--875, July 2008.

\bibitem{ko08p}
A.~Kohnert and S.~Kurz.
\newblock Construction of large constant dimension codes with a prescribed
  minimum distance.
\newblock In Jacques Calmet, Willi Geiselmann, and J\"orn M\"uller-Quade,
  editors, {\em MMICS}, volume 5393 of {\em Lecture Notes in Computer Science},
  pages 31--42. Springer, 2008.

\bibitem{ko08}
R.~K\"otter and F.R. Kschischang.
\newblock Coding for errors and erasures in random network coding.
\newblock {\em Information Theory, IEEE Transactions on}, 54(8):3579--3591,
  August 2008.

\bibitem{ma08p}
F.~Manganiello, E.~Gorla, and J.~Rosenthal.
\newblock Spread codes and spread decoding in network coding.
\newblock In {\em Proceedings of the 2008 IEEE International Symposium on
  Information Theory}, pages 851--855, Toronto, Canada, 2008.

\bibitem{si08a}
D.~Silva, F.R. Kschischang, and R.~Koetter.
\newblock A rank-metric approach to error control in random network coding.
\newblock {\em Information Theory, IEEE Transactions on}, 54(9):3951--3967,
  Sept. 2008.

\end{thebibliography}
\bibliographystyle{plain}

\end{document}